# Single-Molecule Lamellar Hydrogels from Bolaform Microbial Glucolipids

Ghazi Ben Messaoud,[1,†] Patrick Le Griel,[1] Sylvain Prévost,[2] Daniel Hermida-Merino,[3] Wim Soetaert,[4,5] Sophie L. K. W. Roelants,[4,5] Christian V. Stevens,[6] Niki Baccile[1,*]

[1] Sorbonne Université, Centre National de la Recherche Scientifique, Laboratoire de Chimie de la Matière Condensée de Paris, LCMCP, F-75005 Paris, France

[2] Institut Laue-Langevin - 71 avenue des Martyrs CS 20156, 38042 Grenoble, France

[3] Netherlands Organisation for Scientific Research (NWO), DUBBLE@ESRF BP CS40220, 38043 Grenoble, France

[4] Ghent University, Centre for Industrial Biotechnology and Biocatalysis (InBio.be), Coupure Links 653, Ghent, Oost-Vlaanderen, BE 9000

[5] Bio Base Europe Pilot Plant, Rodenhuizekaai 1, Ghent, Oost-Vlaanderen, BE 9000

[6] SynBioC, Department of Green Chemistry and Technology, Ghent University, Coupure Links 653, 9000 Ghent, Belgium.

*Correspondence to: Dr. Niki Baccile, niki.baccile@sorbonne-universite.fr, Phone: 00 33 1 44 27 56 77

† Current address: DWI- Leibniz Institute for Interactive Materials, Forckenbeckstrasse 50, 52056, Aachen, Germany

**Abstract**

Lipid lamellar hydrogels are rare soft fluids composed of a phospholipid lamellar phase instead of fibrillar networks. The mechanical properties of these materials are controlled by defects, induced by local accumulation of a polymer or surfactant in a classical lipid bilayer. Herein we report a new class of lipid lamellar hydrogels composed of one single bolaform glycosylated lipid obtained by fermentation. The lipid is self-organized into flat interdigitated membranes, stabilized by electrostatic repulsive forces and stacked in micrometer-sized lamellar domains. The defects in the membranes and the interconnection of the lamellar domains are responsible, from the nano- to the micrometer scales, for the elastic properties of the hydrogels. The lamellar structure is probed by combining small angle x-ray and neutron scattering (SAXS, SANS), the defect-rich lamellar domains are visualized by polarized light microscopy while the elastic properties are studied by oscillatory rheology. The latter show that both storage G´ and loss G´´ moduli scale as a weak power-law of the frequency, that can be fitted with fractional rheology models. The hydrogels possess rheo-thinning properties with second-scale recovery. We also show that ionic strength is not only necessary, as one could expect, to control the interactions in the lamellar phase but, most importantly, it directly controls the elastic properties of the lamellar gels.

**Introduction**

2D and 3D soft self-assembled materials, usually obtained from stimuli-responsive peptides, proteins and lipids,[1–4] attract a large interest in the field of nanotechnology, for the increasing number of high-tech applications[5] such as protective coating for cells,[6] regenerative medicine,[7] lab-on-a-membrane prototyping,[8] self-healing materials.[9] Lipids can self-assemble into a variety of soft structures,[10] possibly leading to isotropic (entangled fibers) or anisotropic (lamellar) structures, the latter being of particular interest.[11]

Lamellar hydrogels,[12] discovered in 1996 by Safinya and Davidson and composed of a phospholipid $L_\alpha$ phase stabilized by a polymer-grafted lipid, were the first example of an elastic 2D self-assembled material at small concentration (<10 wt%). Since then, lamellar hydrogels (LH) were obtained by polymer-stabilization,[13] or by combining a lamellar phase with a gelator.[14,15] The first polymer-free LH, based on surfactant mixtures or lipid/surfactants, are reported only in 2014.[16,17] Nonetheless, LH are rare compared to more common self-assembled fibrillar network, and their out-of-equilibrium behaviour is unknown. If LH are complex elastic fluids generated by defects,[12] and of which the mechanical properties are hard to control, their defectuous nature is also an opportunity in preparing new materials.[18]

We have recently shown that a pH-responsive C18:0 glucolipid (17-L-[(β-D-glucopyranosyl)oxy]-cis-9-octadecenoic acid, G-C18:0, **Figure 1a**) below 1 wt% undergoes a reversible phase transition from a predominant micellar phase to flat interdigitated lipid layers (**Figure 1b,c**) from basic to neutral pH (transition pH ~7.8) at room temperature (RT).[19,20] The corresponding monounsaturated glucolipid can be classified as a biosurfactant, a family of biobased molecules exclusively produced by the fermentation process of glucose and vegetable oil.[21,22] More specifically, G-C18:0 is obtained by the fermentation of the yeast *S. bombicola ΔugtB1*,[23] and lastly hydrogenated. The microorganism was specially engineered from the WT *S. bombicola*, known to produce the common sophorolipid biosurfactant.[24] Deletion of the second glucosyltransferase (ugtb1) results in the direct production (as high as ~0.5 gL$^{-1}$h$^{-1}$) of glucolipids, up to then produced only by enzymatic conversion of acidic sophorolipids or by the microbial conversion of secondary alcohols glucosides.[25,26] The acidic glucolipid, simple in structure, combines a packing parameter >0.5 with a melting temperature ($T_m$) above RT, thus favouring the spontaneous formation of infinitely flat lamellae over vesicles.[20]

We show here that a phospholipid-free solution only containing glucolipid G-C18:0 (**Figure 1**) in the form of interdigitated membrane spontaneously forms a hydrogel in the neutral-acidic pH range above 1 wt%. The hydrogel is not fibrous, as expected from other microbial glycolipids,[27,28] but it falls in the rare category of LH, without the use of additives like surfactants, polymers or cross-linkers, as shown for other systems.[11] The gels reach elastic moduli above the kPa range, controllable by ionic strength, in a pseudo two-phase lipid-water system. However, pH controls the carboxylic/carboxylate ratio,[19,20] thus hiding an actual four-components (neutral and charged glucolipid, water, salt) system, becoming an unprecedented single-lipid lamellar hydrogel, of which the interactions and, above all, the resulting elastic properties are unexpectedly triggered by ionic strength. The hydrogel forms through the organization of the lipid interdigitated membrane into flat and highly defectuous lamellar regions, having shear-thinning properties with instantaneous (seconds) recovery of the elasticity. The viscoelastic properties highlight a weak power-law frequency dependence of the storage G´ and loss G´´ moduli, a rheological feature of widespread occurrence, but for which existent rheological models failed to establish a clear connection with the microstructure. The straightforward preparation method, in addition to the multi-scale characterization of the system, make this glycolipid lamellar hydrogel a promising candidate for future fundamental and applicative investigations. Finally, the glycosylated nature of the molecule defines a new class of lamellar hydrogel with potential interest in the biomedical field, cosmetics or food science.

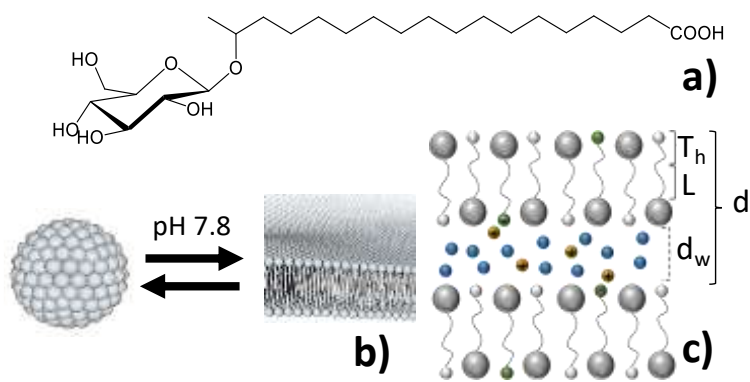

**Figure 1** – a) Molecular structure of glucolipid G-C18:0 (17-L-[(β-D-glucopyranosyl)oxy]-cis-9-octadecenoic acid) and b) its pH-dependent phase behavior at room temperature: a micelles-rich phase occurs at pH above 7.8 and a lamellar phase forms at pH below 7.8.[19,20] c) The lamellar phase is composed of an interdigitated membrane composed of a hydrophilic region ($T_h$), a hydrophobic layer (L) and a water region of thickness $d_w$.

**Materials and Methods**

*Chemicals.* The microbial monunstaurated glucolipid G-C18:1 has been produced at a production rate of ~ 0.5 gL$^{-1}$h$^{-1}$ in a bioreactor system using a modified strain (*ΔugtB1*) of the yeast *Starmerella bombicola*[23] and according to the experimental conditions described in ref.

[20]. The fully saturated G-C18:0 ($M_w$ = 462.6 g.mol$^{-1}$), used in this work, was obtained from GC18:1 by a catalytic hydrogenation reaction, described in ref. [20]. The NMR, HPLC and LC-MS analyses of G-C18:0 can be found in ref. [20]. 18:1 Liss Rhod PE ($M_w$= 1301.7 g.mol$^{-1}$, $\lambda_{abs}$= 560 nm, $\lambda_{em}$= 583 nm), 1,2-dioleoyl-sn-glycero-3-phosphoethanolamine-N-(lissamine rhodamine B sulfonyl) (ammonium salt), is purchased by Avanti® Polar, Inc

*Preparation of G-C18:0 hydrogels.* Hydrogels are prepared by dispersing the G-C18:0 powder in water followed by sonication and adjustment of pH to the desired pH value and ionic strength. In detail, a given amount of G-C18:0 (in wt%) is dispersed in a given volume of milliQ water (generally 1 mL). The pH of the mixture is generally between 3.5 and 4.5, according to the sample concentration. The pH is then adjusted in the range 5.5 - 7.5 using 1-20 µL of NaOH 1 M (0.1 M can also be used for refinement), according to concentration and desired pH. The mixture is then sonicated between 15 and 20 min in a classical sonicating bath to reduce the size of the aggregated powder and until obtaining a homogenous, viscous, dispersion. To this solution, the desired volume of NaCl is added so to obtain a given total [Na$^+$] {= [NaOH] + [NaCl]} molar concentration. To keep the dilution factor negligible, we have used a 5 M concentrated solution of NaCl. The mixture is then sonicated again during 15 min to 20 min and eventually vortexed two or three times during 15 s each. The solution can then be left at rest during 15 min to 30 min. This procedure is generally enough to obtain a stable hydrogel. However, to reduce the impact of gel history, rheological measurements are performed after 2 hours following an annealing cycle: vortexing of the gel during 15 s at room temperature, loading of the fluid solution in the rheometer, heating from 25°C to 70°C (above $T_m$) at 10°C/min, rest at T= 70°C during 10 min, cooling to T= 25°C at 10°C/min.

*Small Angle X-ray Scattering (SAXS).* All SAXS experiments are performed at 25°C at the DUBBLE BM26B beamline at the ESRF synchrotron facility (Grenoble, France).[29,30] Samples have been analysed during the run SC4639 using a beam at 11.93 KeV and a sample-to-detector distance of 2.10 m. Samples are prepared ex-situ and they are analysed directly by setting them in front of the x-ray beam. The signal of the Pilatus 1M 2D detector (172 x 172 µm pixel size), used to record the data, is integrated azimuthally with PyFAI to obtain the *I(q) vs. q* spectrum ($q = 4\pi \sin\theta / \lambda$, where 2θ is the scattering angle) after masking systematically wrong pixels and the beam stop shadow. Silver behenate ($d_{(100)}$ = 58.38 Å) and α-Al$_2$O$_3$ ($d_{(012)}$ = 3.48 Å) are respectively used as SAXS and WAXS standards to calibrate the q-scale. Data are not scaled to absolute intensity.

*Small Angle Neutron Scattering (SANS).* SANS experiments have been performed at the D11 beamline of Institut Laue Langevin (Grenoble, France). Four q-ranges have been explored and merged using the following wavelengths, λ, and sample-to-detector (StD) distances. 1) ultra-low q: λ= 13.5Å, StD= 39 m; 2) low-q: λ= 5.3Å, StD= 39 m; 3) mid-q: λ= 5.3Å, StD= 8 m; 4) high-q: λ= 5.3Å, StD= 1.4 m. All samples are prepared in 99.9% $D_2O$ (using NaOD and DCl solutions, as well) to limit the incoherent background scattering. Solutions are analyzed in standard 1 mm quartz cells. Direct beam, empty cell, $H_2O$ are recorded and boron carbide (B4C) is used as neutron absorber. All samples are thermalized at 70°C during 10 min then cooled down at 25°C. Analysis is performed after 2 h from cooling. The background sample ($D_2O$) signal was subtracted from the experimental data. Absolute values of the scattering intensity are obtained from the direct determination of the number of neutrons in the incident beam and the detector cell solid angle. The 2-D raw data were corrected for the ambient background and empty cell scattering and normalized to yield an absolute scale (cross section per unit volume) by the neutron flux on the samples. The data were then circularly averaged to yield the 1-D intensity distribution, I(q). The software package Grasp (developed at ILL and available free of charge) is used to integrate the data, while the software package SAXSUtilities (developed at ESRF and available free of charge) is used to merge the data acquired at all configurations and subtract the background.

*Rheology.* Rheology experiments are carried out using a MCR 302 rheometer (Anton Paar, Graz, Austria) equipped with a Peltier temperature system which allows accurate control of the temperature by the stainless steel lower plate and with a solvent trap to ensure minimal evaporation of water during the measurements.
- *Oscillatory rheology*. These experiments are performed using titanium or a stainless steel sandblasted upper plate (diameter 25 mm). The gap (0.5 or 1 mm) and the normal force (NF= 0 N) are controlled during the experiments. After loading, samples are allowed to stand at rest for at least two hours before analysis. A dynamic strain sweep is first conducted at an angular frequency (ω= 6.28 rad·s$^{-1}$) by varying the shear strain (γ) from 0.001 to 100 % in order to determine the linear viscoelastic regime (LVER). A value of strain within the LVER is then applied in the following dynamic angular frequency sweep between 100 and 0.01 rad.s$^{-1}$. To monitor the gelation of the G-C18:0 upon thermal annealing, the elastic (G´) and viscous (G´´) moduli are recorded during temperature heating ramps from 20 to 70 °C at a rate of 10 °C/min. The sample is initially vortexed, then loaded and held at 70 °C for 10 min and then cooled from 70 to 25 °C at a rate of 10 °C/min and finally held at 25°C during two hours. These temperature variation experiments are performed using an oscillation angular frequency ($\omega$= 6.28 rad·s$^{-1}$) and a strain ($\gamma$ = 0.05%). Afterward, an angular frequency sweep (100 – 0.01 rad·s$^{-1}$) was performed using a shear strain, $\gamma$= 0.05%, within the LVER.

- *Shear viscosity*. Steady-shear viscosity is determined using a cone-and-plate geometry (diameter 25 mm, angle 1° and truncation 50 μm) and a plate-and-plate geometry (diameter 25 mm, gap = 0.5 mm) by increasing the shear rate ($\gamma$) from 10$^{-3}$ to 10$^{3}$ s$^{-1}$. Compared to plate-plate geometry, cone-plate configuration guarantee a uniform shear of the sample, however the small imposed gap (truncation 50 μm) could lead to preferential aggregation of the lamellar structure (3-4 nm of thickness but few μm of length (L$_{lamellar}$) in the center of the geometry (L$_{lamellar}$ > gap x 10). Here we presented therefore the data obtained used the plate-plate geometry. Except for thermal annealing experiments, all rheological characterizations are conducted at 25 °C, unless otherwise mentioned.

*Light Microscopy.* Images of G-C18:0 samples, at rest or after shear, are acquired in bright field and polarized light mode (PLM) using a Nikon DS-Ri1 through crossed polarizers (samples are prepared in flame-sealed flat capillary of 200 μm thickness) and in a differential interference contrast mode (DIC-M) using a Zeiss AxioImager D1 microscope.

*Differential Scanning Calorimetry (DSC)*: DSC is performed using a DSC Q20 apparatus from TA Instruments equipped with the Advantage for Q Series Version acquisition software (v5.4.0). Acquisition is performed on a G-C18:0 dry powder sample (~ 10 mg) sealed in a classical aluminium cup and using an immediate sequence of heating (from 10°C to 90°C) and cooling (from 90°C to 10°C) ramps both at a rate of 1°C/min.

*Confocal Laser Scanning Microscopy (CLSM)*: CLSM was performed with a LeicaSP8 Tandem Confocal system. Samples were excited with the dye specific wavelength (561 nm) and the emission was detected between 580 and 620 nm using a photomultiplier tube (PMT) detector. CLSM images were analyzed using Fiji (Fiji is just ImageJ)[31] and 3D construction was performed using the 3D Stack mode of Fiji. Temperature variation (T= 50°C) was performed with temperature controller modulus of the microscope. The hydrogel ($C_{G-C18:0}$= 2.5 wt%, pH 6) was prepared following the general method described above. A volume of 4 µL of an ethanolic solution of 18:1 Liss Rhod PE (C= 53 mg/mL) was added to 1.5 mL of the hydrogel to reach an approximate molar ratio of G-C18:0/Liss of 500. Liss is a water insoluble, rhodamine-containing, lipid and it is largely used to mark lipid bilayers. It is generally considered not to interfere with the bilayer assembly at Lipid/Liss ratio above 200. We did not observe any variation in the gel physical aspect after addition of Liss.

**Results and discussion**

*Lamellar structure of the glucolipid G-C18:0 probed by SAXS and SANS.*

G-C18:0 is a microbial glucolipid characterized by a single glucose moiety covalently attached to stearic acid, a melting temperature, $T_m$, of about 37°C (**Figure S 1**) and a pKa of 8.4 (**Figure S 2**). A combination of SAXS and cryo-TEM data reported earlier[19,20] have shown the formation of an interdigitated layer (IL) exposing both the glucose and COOH/COO⁻ moieties, where the COOH/COO⁻ ratio depends on pH. **Figure 1** displays a scheme of the supposed structure of a G-C18:0 layer characterized by hydrophilic ($T_h$), hydrophobic (L) and water layers ($d_w$). The total thickness of approximately 3.6 nm ($2T_h + L$), estimated after modelling SAXS profiles using a standard bilayer model,[19,20] is compatible with the full length of G-C18:0, thus suggesting the stabilization of an interdigitated membrane rather than a bilayer, classically found for phospholipid membranes. The typical cryo-TEM images, extensively presented elsewhere,[19,20] show the presence of infinitely wide G-C18:0 stacked layers.

SANS (**Figure 2a-c**) and SAXS (**Figure 2d**) experiments are recorded for samples at pH 7 and pH 6.5 and various NaCl concentrations and concentrations between 1 wt% and 5 wt%. SAXS profiles measured at NaCl concentration below 250 mM bear a large oscillation above q= 1 nm⁻¹ (**Figure 2d**) and a -2 slope (in log-log scale) below q= 1 nm⁻¹. The latter is typical for 2D objects and in particular for lipid membranes,[32] while the latter is a typical feature for bilayers and previously found in the same G-C18:0 systems and fitted using a lamellar form factor model.[19,20] One should note that the different scattering length densities of G-C18:0 and water with respect to x-rays and neutrons induces an important difference in terms of the contrast between the lipid and water, and for this reason the oscillation above q= 1 nm⁻¹ is not observed in SANS profiles. Nonetheless, both SAXS and SANS profiles are characterized by a broad peak below 1 nm$^{-1}$ and indicative or a lamellar period, d$_{(100)}$. Whichever the pH or technique of analysis, the peak is prominent for G-C18:0 concentrations above 2.5 wt% and NaCl content above about 100 mM, while below these values the peak is extremely broad and hard to observe. To better observe the first and second order of the lamellar period, one can eliminate the q$^2$ dependence by mean of Kratky plots (Iq$^2$(q)); **Figure 2a-c** show the Kratky plots corresponding to all samples, thus putting in evidence, for some of them, the presence of both the d$_{(100)}$ and d$_{(200)}$ reflections, where the latter are unfortunately too broad to be observed in most cases. At high ionic strength for volume fractions above 2.5 wt%, one can observe a second broad peak at q-values below about 0.7 nm$^{-1}$. Attribution of this peak, of which we do not observe the second order, is not straightforward but we make the hypothesis of the coexistance of two lamellar domains stabilized by water layers of different thickness, probably induced by a non-homogeneous and specific adsorption phenomena of Na$^+$ onto the membrane, as described theoretically and found experimentally in charged lipid membranes.[33–35]

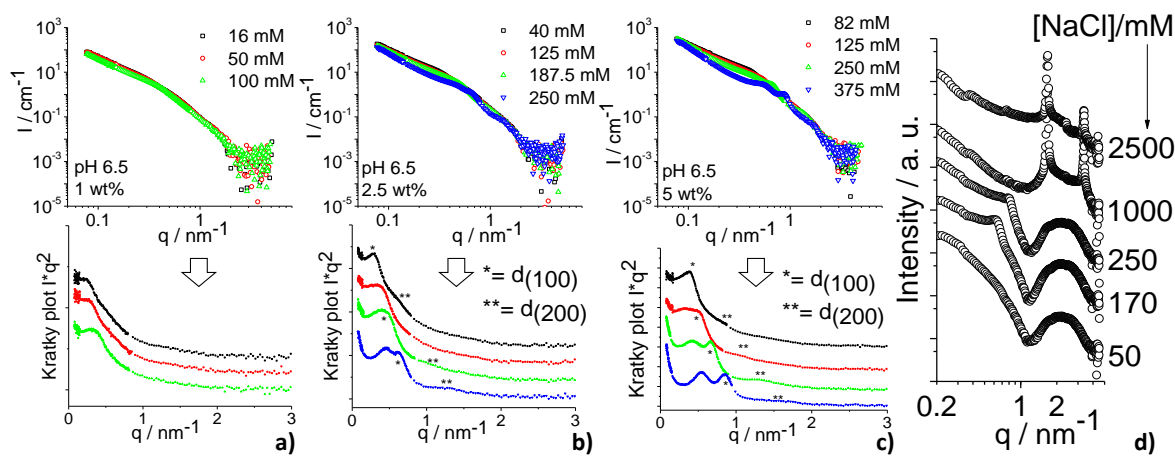

**Figure 2 - a-c) SANS profiles acquired on samples prepared at pH 6.5 and various NaCl concentrations at (a) 1 wt%, (b) 2.5 wt%, (c) 5 wt%. Kratky plots [I·q$^2$(q)] are provided below each series of spectra. d) SAXS profiles acquired for a G-C18:0 solution (5 wt%) at pH 7 at various NaCl concentrations.**

Analysis of d$_{(100)}$ for all SAXS and SANS experiments shows that d$_{(100)}$ decreases with increasing ionic strength (**Figure 3a**), an evidence which is commonly observed for lipid lamellar phases stabilized by electrostatic repulsive interactions.[36] In the excess of salt (1 M), a powder, characterized of condensed lamellae, precipitates, as also shown by diffraction peaks becoming sharper, more intense, and shifting above 1 nm$^{-1}$ (**Figure 2d).** The expected period for an ideally swelled lamellar phase is d$_{ideal}$=δ/φ, being δ the membrane thickness (here assumed to be 2T$_h$ + L = 3.6 nm)[20] and φ the lipid volume fraction. Plotting the experimental d-spacing over $d_{ideal}$, $d/d_{ideal}$, against the glucolipid volume fraction, $\varphi_{GC}$, one systematically finds $d/d_{ideal}<1$ (**Figure 3b**), meaning that swelling does not follow ideality below 10 wt% at any pH and ionic strengths. Similar results were obtained for diluted phospholipid lamellar phases below 20 wt% and explained by the presence of a second phase applying an additional osmotic pressure onto the lamellae.[37] Similar arguments could certainly explain $d/d_{ideal}<1$ in the present system.

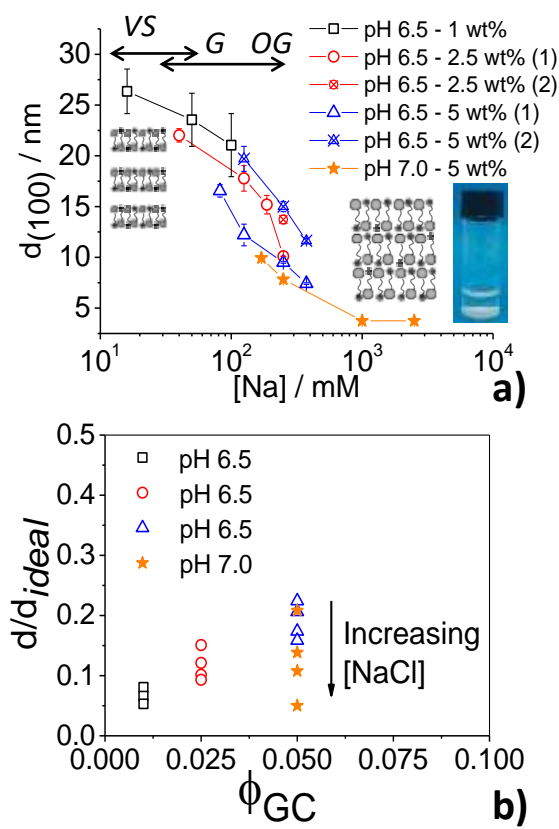

**Figure 3 - a) Evolution of the lamellar period $d_{(100)}$ as a function of ionic strength; the corresponding SAXS and SANS data are shown in Figure 2. *VS*, *G* and *OG* correspond to macroscopic state of the sample: *VS*, Viscous solution; *G*, Gel; *OG*, Opal Gel. b) $d/d_{ideal}$ ratio as a function of G-C18:0 volume fraction for different ionic strengths and pH values; $d_{ideal}= \delta/\varphi$, being δ the membrane thickness (= 3.6 nm)[20] and φ the lipid volume fraction.**

Attribution of the lipid packing within the membrane and type of lamellar phase is done using the WAXS portion of the x-ray scattering data (**Figure 4a**). The WAXS data show a sharp peak at 14.9 nm$^{-1}$: in the absence of salt, the peak is so broad that it can hardly be detected, unless background (water, characterized by the classical peaks at 20 and 27 nm$^{-1}$) is subtracted (**Figure 4b**). Above [NaCl]= 1 M, the peak is intense and sharp. All experiments are performed at room temperature, below the $T_m$ (**Figure S 1**) thus suggesting an ordered packing of the lipids within the membrane. It is then expected that G-C18:0 forms a $L_{\beta,i}$ phase (interdigitation comes from the bolaform nature of G-C18:0). A classical $L_\beta$ is commonly characterized by an intense, well-defined, peak between 14.9 nm$^{-1}$ and 15 nm$^{-1}$ ($d$-spacing of 0.42 nm),[38–40] signature of the parallel packing of the acyl chain within the hydrophobic region of a bilayer. The collapsed phase is certainly in the form of $L_{\beta,i}$. On the contrary, in the swollen region, the same peak is very broad; according to the water-containing profiles (b), the shoulder at 14.9 nm$^{-1}$ should be attributed to a $L_\alpha$ phase, characterized by a liquid-like order of the acyl chain. However, the expected $q_0$ should rather be at 14.7 nm$^{-1}$ ($d$-spacing of 0.46 nm),[38–40] a smaller value than what we do observe experimentally. The last possibility, probably the most plausible, is represented by the $P_\beta$ phase, which is characterized by a single peak with a corresponding $d$-spacing around 0.42 nm and a width larger than the $L_\beta$ phase (although not as large as in the $L_\alpha$).[38,40]

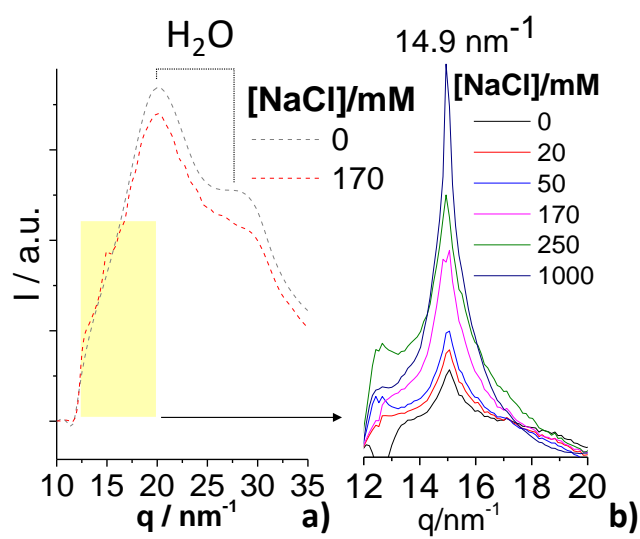

**Figure 4 - a) Background (water)-containing and b) background-subtracted WAXS profiles of samples studied in Figure 2d.**

*Glucolipid G-C18:0 spontaneously forms lamellar hydrogels in water without additives*

G-C18:0 hydrogels can be formed by dispersion of the lipid in water at room temperature, at concentrations above about 1 wt%. Gel formation is very straightforward as long as the following procedure is employed. The equilibrium pH of G-C18:0 is generally below 4.5, under which G-C18:0 is insoluble in water. Hydrogels can then be obtained by combining adjustment of pH in the range between 5 and 7.5 with homogenization during 20 to 30 min in a sonication bath. These conditions generally provide a viscous solution (*VS*, **Figure 5a**) with a corresponding d-spacing of ~20 nm (*VS* in **Figure 3a**), which turn into a gel when salt (here, NaCl) is added to concentrations above 20 mM. According to the amount of salt, hydrogels look homogeneous and do not recover immediately after vortexing (**Figure 5b**), or they can display an opal-like appearance and fast recovery (**Figure 5c**). The corresponding d-spacing for a gel (*G*) and opal gel (*OG*) are generally below 15 ~nm (*G* and *OG* in **Figure 3a**).

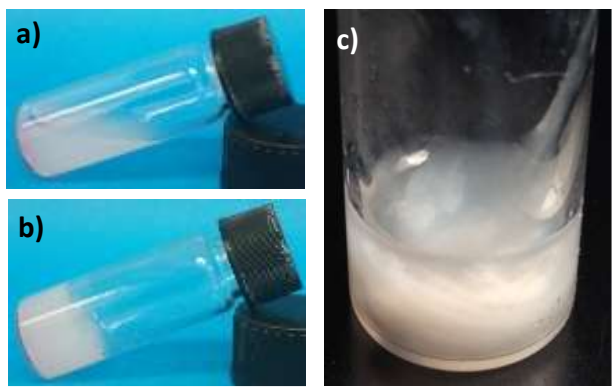

**Figure 5** –Typical images of (a) viscous (V), (b) gel (*G*) and c) opal gel (*OG*) solutions. *OG* in c) is denser and presents a stronger scattering than gel (*G*) in b).

In order to perform comparable rheology experiments, all samples studied in **Figure 5** were annealed through vortexing, heating at 70°C, above the $T_m$ (**Figure S 1**) and cooling in the rheometer. The typical sol-to-gel transition upon cooling is measured in **Figure 6a** over 2 h, during which both moduli (G´´ > G´) have a steep increase, then oscillating at 25°C until G´ > G´´. The typical angular frequency sweep (G´(ω) and G´´(ω), respectively represented by full and open symbols throughout this work) of a (annealed) lamellar G-C18:0 hydrogel prepared at pH 6.5 for three ionic strengths are shown in **Figure 6b**. The typical range of G´ (ω= 6.28 rad·s$^{-1}$) spans between $10^2$ Pa and $10^4$ Pa when ionic strength ranges between ~50 mM and ~ 500 mM, measured at pH 6 and pH 6.5 (**Figure 6c**). The magnitude of G´ falls in the same order as observed for polymer-stabilized $L_{α,g}$[41] and polymer-free lamellar hydrogels.[16] **Figure 6b** also shows that G´ > G´´ over the entire angular frequency range, an established criterion distinguishing gels from viscous liquids.[42] The typical viscoelastic response of G-C18:0 follows a power-law behavior over four orders of angular frequency magnitude. Both moduli scale as G´~ $G_0´ω^α$ and G´´~ $G_0´´ω^β$, where $G_0´$ and $G_0´´$ are pre-factors at ω= 1 rad·s$^{-1}$ and $α, β$~ 0.23 the exponents. In general, for stable hydrogels we find the exponents $α$ and β spanning between 0.15 and 0.3, while for unstable hydrogels with higher amount of NaCl (142 mM), the system behaves like a viscoelastic fluid with an elastic character (G´> G´´) at high ω followed by a crossover point (G´= G´´) at intermediate ω and finally (G´< G´´) at low ω.

Time stability of the hydrogels over 24 h shows that weak gels ($10^2$ Pa) measured after 2 h from annealing (blue symbols in **Figure 6d**) evolve towards stronger gels ($10^3$ Pa) after 24 h, indicating that the evolution of the mechanical properties can occur on long time-scales.

Quantitative observations up to 48 h (**Figure S 3**), and qualitatively over two weeks, show that initially viscous solutions systematically turn into gels, which eventually tend to scatter light (whitening) and to show syneresis. The continuous evolution of the aspect and mechanical properties show that none of the lamellar hydrogels reach thermodynamic equilibrium, which makes it impossible to draw a consistent quantitative overview correlating mechanical properties and physico-chemical parameters. For this reason, experiments presented in **Figure 6a,b** are performed after an arbitrary lag time of two hours, at which the estimated concentration above which gelation is possible (here measured at pH = 6.7, ionic force 163 mM, **Figure S 4**) is in the order of 1 wt%.

pH controls the ionization degree of G-C18:0 and, as a consequence, on the charge density of the lipid membranes. To this regard, it is well-known that ionic strength strongly affect the interbilayer distance in charged lamellar systems.[33,34,43–45] but the impact of salt on the mechanical properties of lamellar hydrogels is particularly unexpected, expecially at low lipid volume fractions. It was never reported for lamellar gels and only partially investigated in concentrated (> 10 wt%) onion phases, with uncomparable impact on the elastic properties (G´< 10 Pa for concentrations in the order of 10 wt%)[43] with respect to the system shown here. The elastic properties of the G-C18: lamellar hydrogel are extremely sensitive to ionic strength above about 20 mM to 50 mM. At ionic strength below ~100 mM, pH plays a marginal role on the elastic properties. However, at higher ionic strength, charge screening effects induce the collapse of the membrane (**Figure 3a**) and which can result in macroscopic phase separation (precipitation). If pH is increased, additional negative charges are most likely introduced in the glucolipid membrane following deprotonation of the carboxylic acid and recovery of the gel is possible, possibly due to swelling. On the contrary, at small salt concentrations and constant pH between 5 and 7, the solution is generally viscous; by increasing the ionic strength below molar amounts, gel formation is generally promoted, a phenomenon systematically associated to lamellar shrinking, with inter-layer water thicknesses, $d_w$, dropping from 25-20 nm to about 15 nm (**Figure 3a**).

Finally, **Figure S 4** and **Figure S 5** show the rheo-thinning and thixotropic properties of G-C18:0 hydrogels under high shear and high strain conditions. In particular, a typical hydrogel (2.5 wt%, pH 6, [NaCl]= 123 mM) submitted to a series of step-strain (0.5 % < $\gamma$ < 100 %) cycles is able to recover about 25% of its initial elasticity after 30 s and between 60% and 80% after 30 min (**Figure S 5**).

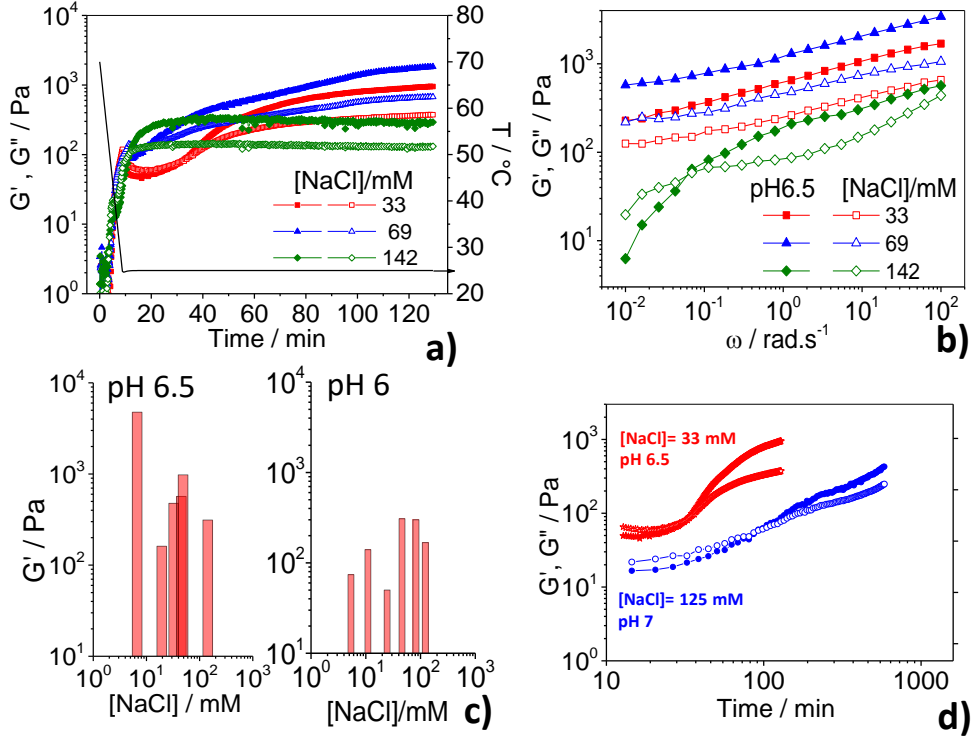

**Figure 6** – a) Time evolution of G´ (full symbols) and G´´ (open symbols) at 25°C for various ionic strengths (cooling rate: 5°C/min, $C_{G-C18:0}$= 2.5 wt%, pH= 6.5, ω= 6.28 rad·s$^{-1}$, γ= 0.05%) after annealing (15 s vortex, loading, T= 70°C during 10 min, heating rate: 10°C/min). b) G´(ω) (full symbols) and G´´(ω) (open symbols) for $C_{G-C18:0}$= 2.5 wt% as a function of [NaCl] (pH= 6.5, γ= 0.05%). c) Typical absolute values of G´ at pH 6 and 6.5 as a function of [NaCl] (ω= 6.28 rad·s$^{-1}$, γ= 0.05%). d) Comparison of the time evolution of G´ (full symbols) and G´´ (open symbols) after annealing at T= 70°C and cooling at 25°C for a gel ([NaCl]= 33 mM, pH 6.5) and a viscous solution ([NaCl]= 125 mM, pH 7) (cooling rate: 5°C/min, $C_{G-C18:0}$= 2.5 wt%, ω= 6.28 rad·s$^{-1}$, γ= 0.05%). All rheology measurements are performed in a plate-plate geometry at constant normal force of 0 N (initial gap 0.5 mm).

*Understanding the microstructure of G-C18:0 hydrogels*

From a rheology point of view, the distinctive power law response is commonly reported for several soft materials like cells[46], adipose tissue[47], collagen[48], colloidal[49,50] and polymeric gels[51–53] and indicates a broad range of relaxation times. Power law viscoelasticity is generally captured using three distinct models namely critical gels,[51] soft glassy rheology (SGR)[54] and fractional rheological models.[55] In the current case, the critical gel (system at the gel point) model is not suitable to describe the rheological response of G-C18:0 hydrogels since G´(ω) and G´´(ω) have to remain strictly parallel with the frequency resulting in a frequency independent loss factor (tan (δ) = G´´(ω)/G´(ω)). This is not the case since in many experiments we observed G´(ω) ~ $ω^α$ and G´´(ω) ~ $ω^β$ with α ≠ β. Moreover, at high salt concentrations, a power-law behavior over high frequency range was observed but with a G´(ω) ~ G´´(ω) crossover at low frequencies (**Figure 6b**). The SGR model considers that viscoelasticity is controlled by disorder, metastability and local structural rearrangements between the mesoscopic elements.[56] Thermal motion alone is not sufficient to reach complete relaxation and the system has to cross energy barriers, larger than typical thermal energies, and related to lamellar rearrangement in our case. The system can be driven into a glassy, liquid, or intermediate state through a mechanically-activated process characterized by a mean-field ''noise temperature'', $x$. G-C18:0 hydrogels satisfy all criteria of a soft glassy material: storage moduli magnitude (G´~ 0.1-10 kPa), aging behavior and weak power law of G´, G´´(ω). Compared to canonical rheological models based on the combination of springs and dashpots, fractional rheological models are based on the use of spring-pots elements derived using fractional calculus. The spring-pot element is characterized by a fractional exponent ($\alpha$) with $0 < \alpha < 1$, where for $\alpha \rightarrow 1$ and $\alpha \rightarrow 0$, the fractional element is assumed as dashpot and spring, respectively.[57] Both the SGR and fractional rheological models based on the combination of spring-pot elements are suitable for capturing the power-law viscoelasticity. Jaishankar and McKinley had even demonstrated that the constitutive equation of the SGR model can be reduced to that of a single spring-pot element by relating the fractional exponent ($\alpha$) to the noise temperature ($x$) by $\alpha = x-1$.[57]

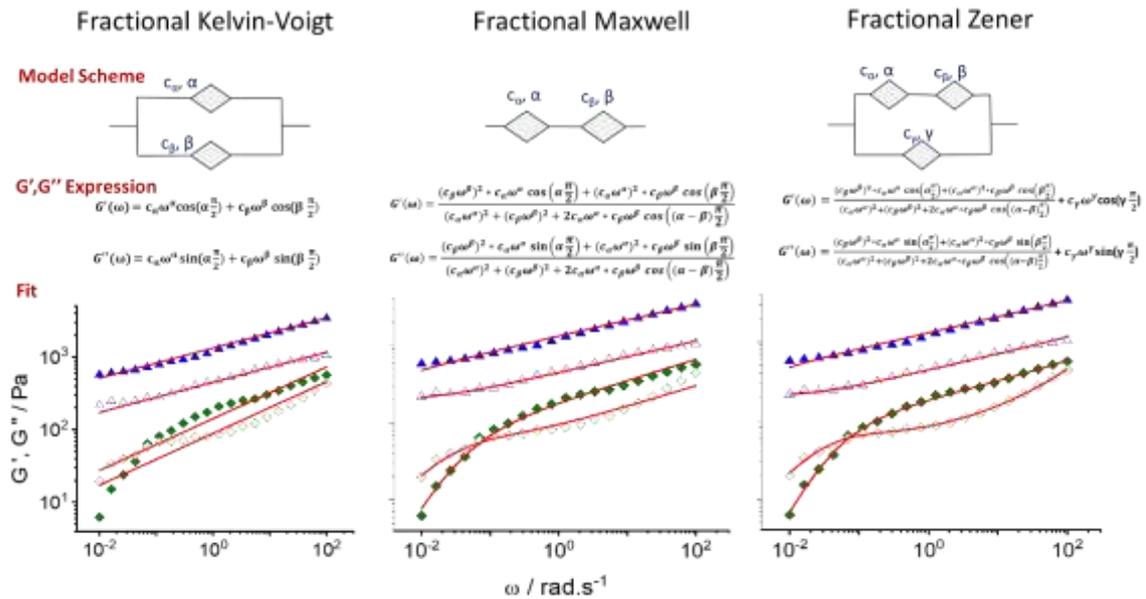

**Figure 7 – G´(ω) (full symbols) and G´´(ω) (open symbols) for $C_{G-C18:0}$= 2.5 wt% as a function of [NaCl] (69 mM, blue; 142 mM, green) at pH= 6.5 and γ= 0.05%. The data are fitted using three fractional models:**

**Kelvin-Voigt (2 parallel spring-pots), Maxwell (series of 2 spring-pots) and Zener (combination of both). The fitting procedure was achieved using the open source library RHEOS**[58] **developed in Julia language.**[59]

Based on that, we attempt to model the viscoelastic response of G-C18:0 hydrogels by fitting the rheological data presented in **Figure 6b** to a combination of two spring-pot elements in series (fractional Maxwell) and in parallel (fractional Kelvin-Voigt) and to a combination of three spring-pots (fractional Zener) (**Figure 7**). The parameters resulting from the fitting procedure are shown in Table 1. The best fit was obtained using the fractional Zener model. The fractional element (β) demonstrates that the value of β increases from 0.18 to 0.22 upon NaCl addition but the most significant change is observed for the fractional element γ with γ =0.25 (<0.5: a dominant spring character) to γ =0.83 (>0.5: a dominant dashpot character) confirming the loss of mechanical properties and in a good agreement with the observed crossover at low frequency (from hydrogel to a viscoelastic fluid). For both hydrogels the $\alpha$ =0.99 and therefore this fractional element could be substituted by a simple dashpot element ($\alpha$ ~1). An optimum model combined a dashpot and two spring-pots elements could be therefore envisaged.

**Table 1. Fitting parameters from the curves of Figure 7.**

| Sample | Fractional model | Parameters | | | | | |
|---|---|---|---|---|---|---|---|
| [NaCl] | | $C_\alpha$ | $\alpha$ | $C_\beta$ | $\beta$ | $C_\gamma$ | $\gamma$ |
| **69 mM** | Fract Kelvin-Voigt | 0.0 | 0.95 | 1406.8 | 0.21 | | |
| | Fract Maxwell | 1394.29 | 0.21 | 475669.9 | 0.99 | | |
| | Fract Zener | 260141.31 | 0.99 | 993.21 | 0.18 | 397.31 | 0.25 |
| **142 mM** | Fract Kelvin-Voigt | 167.29 | 0.36 | 0.0 | 0.22 | | |
| | Fract Maxwell | 208.09 | 0.27 | 2713.19 | 0.99 | | |
| | Fract Zener | 2517.02 | 0.99 | 201.07 | 0.22 | 6.13 | 0.83 |

Based on the sum of squared residual errors, the fractional Zener model provide the best fit followed by the fractional Maxwell and lastly the fractional Kelvin-Voigt which failed to fit the behavior of G-C18:0 hydrogel at a high NaCl concentration (142 mM at pH6.5). Even if a direct relationship between microstructure and mechanical response remains challenging,[60] the power law viscoelasticity highlighted by the broad range of relaxation times is related to the presence of hydrogel heterogeneities: structures of different sizes (interdigitated layer with different lateral size, defects and curvatures, spherulites domains) and densities (accumulation of counter-ions in some regions more than in others). Moreover, the evolution of the rheological response with time and salt concentration combined to the microstructural characterization point out that the mechanical properties of G-C18:0 are mainly driven by inter-lamellar interfaces. Such analysis is consistent with the optical microscopy data presented here below and coherent with what is known in other 2D materials like graphene oxide, which only displays a power-law rheology behaviour when the graphene layers are strongly interacting after addition of a polymer matrix.[61]

**Figure 8a,b** (additional images are shown in **Figure S 6g-h**) puts in evidence the strong heterogeneity of G-C18:0 gels. 1-labeled arrows show oriented defectuous lamellar domains both in bright field (**Figure 8a**) and under crossed polarizers (**Figure 8b**). These domains look like the whispy textures in Ref. 62 and they are connected to defectuous birefringent, although disordered, domains (arrows N° 3) through a continuous change in orientation (arrows N° 2). Finally, 4-labelled arrows point at spurious small (< 10 μm) spherulitic domains. The entire set of polarized light microscopy experiments show that the G-C18:0 lamellar hydrogel is not composed of a simple defectuous single lamellar phase, as classically found in lamellar hydrogels,[12,41] but it is rather composed of large (hundreds of microns) interconnected defectuous lamellar domains dispersed in water. For sake of comparison, **Figure S 6** presents a broad series of bright field and polarized light microscopy (indicated with the *CP*, Crossed Polarizers, subscript) images recorded on typical fluid and viscous samples composed of G-C18:0 at pH 6.5. Liquid samples are generally characterized by small strongly birefringent circular domains of approximate size of 1-10 μm at 1 wt% and room temperature (**Figure S 6a-b**) and 10-50 μm at 2.5 wt% (30 mM NaCl) at T= 60°C (**Figure S 6e-f**). These inclusions recall the spherulites in ref. 41 and they are dispersed in a non-birefringent medium. Nevertheless, the nature of these domains is most likely not the same; temperature is known to induce a lamellar-to-vesicle transition,[20] while the 10-50 μm spherulitic domains in the sample at T= 60°C (2.5 wt%, 30 mM NaCl, **Figure S 6e-f**) correspond to vesicular, probably multilamellar, compartments, given their strong birefringency. Upon increasing in viscosity (2.5 wt%, 30 mM NaCl, RT, **Figure S 6c-d**), the medium becomes more and more characterized by larger irregular birefringent domains of whispy texture (typical size of 100-500 μm) coexisting with strongly birefringent spherulitic inclusions of smaller size (1-50 μm), as discussed above.

A global microstructural description of the gels can then be summarized as follows: 1) large (100-500 μm) oriented domains recalling the typical sheet-like/whispy texture of lamellar hydrogels (Figure 17G in Ref. 62); 2) an aqueous matrix composed of interconnected lamellar domains below 5 μm in size; 3) embedded spherulitic objects composed of folded lamellar domains of thickness less than 500 nm. The latter are widely present in the gel and they are themselves composed of entangled lamellar domains. **Video 1,2** and **Video 3** respectively confirm the presence of domains below 5 µm and spherulitic inclusions. Finally, microscopy experiments confirm the presence of two phases and which can explain the non-ideal swelling of the lamellar phase against $\varphi_{GC}$ (**Figure 3b**).

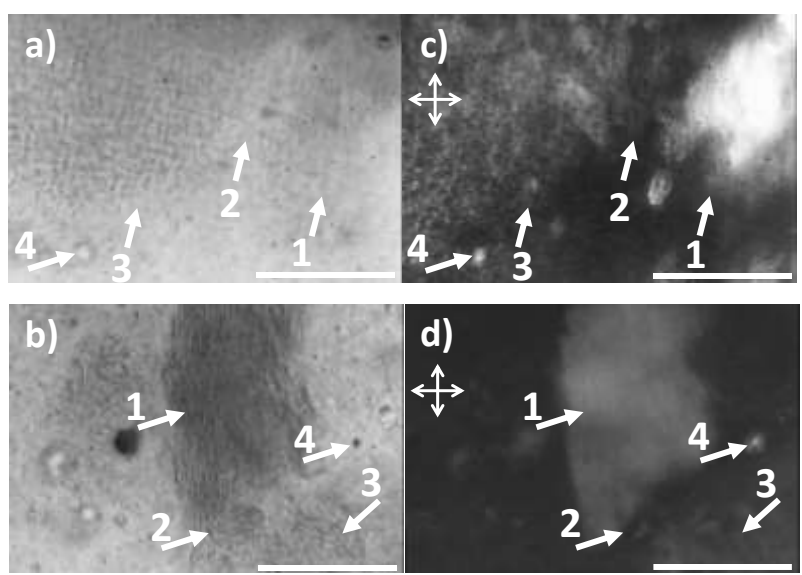

**Figure 8 – a,b) Bright field and c,d) polarized light microscopy experiments (scale bar: 100 µm) experiments recorded on G-C18:0 hydrogel at C= 2.5 wt%, pH 6.5 and [NaCl]= 150 mM. Arrows 1 through 3 correspond to anisotropic regions with different orientations and arrow 4 to spherulitic domains.**

*Origin of the elastic properties of G-C18:0 lamellar hydrogels*

The elastic properties of lamellar hydrogels are classically explained by the presence of structural defects in a monophasic lamellar system. Gel formation in polyethyleneglycol(PEG)-stabilized $L_\alpha$ phase depends on the PEG volume fraction, which, segregating into curved membrane regions, become a nucleation site of the defects.[12,41] Recently, Niu et al.[16] have supposed the origin of the elastic properties in polymer-free LH to depend on the edge accumulation of negatively-charged SDS surfactant. To this regard, polarized light microscopy (PLM) is commonly used to prove the presence of structural defects, like sheet-like texture or spherulitic inclusions, in lamellar hydrogels.[12,62] The morphology of the lamellar domains and type of defects is very important to identify the nature of the phase and, consequently, the relationship to the macroscopic elastic properties.

In the present system, XRD and PLM experiments show non-ideal swelling, typical of lipid lamellar phases,[36,37,63] and demonstrate that the system is biphasic. PLM also demonstrates the presence of highly defectuous lamellar domains and it excludes the existance of fibrillar crystals, commonly responsible for the elastic properties in coagels.[64] The absence of oily streaks and Schlieren defects reasonably exclude, respectively, the existence of defectless lamellar and nematic phases (additional comments on these points are given in the Supporting Information).[41,62,63] Despite the high density of defects shown by PLM, their nature strongly recalls the whispy texture shown in lamellar gels,[41,62] and characterized by various types of defects including screw dislocations, dislocations, disclinations or spherulite inclusions, as shown in **Figure 9b**.[12,41,62,63] **Figure 9a,b** finally summarizes the above by showing the micro and submicro structure of the G-C18:0 gels: large, highly defectuous, lamellar domains interconnected to each other.

At a molecular scale, the stabilization of a defect could be explained by the presence of either an end-capped membrane (**Figure 9d**) or a curved membrane (**Figure 9d**). The former can occur if the line tension is small and the latter if the membrane is able to bend. The lamellar phase forms at pH below the pKa, determined here to be above 8 (**Figure S 2**), where the carboxylic form of G-C18:0, although majoritary, still coexists with the negatively-charged carboxylate form (**Figure 9c**). End-capping can be explained by a local accumulation of the carboxylate form of G-C18:0 in the cap (**Figure 9d**). This hypothesis is not outrageous because the curvature of the cap is comparable to the curvature of a micelle, the latter being the stable phase of G-C18:0 in its fully deprotonated form.[19]

Bending of the membrane is the other phenomenon that generates defects and it could be explained as follows. At room temperature, the membrane composed of G-C18:0 is expected to be rigid. However, bending could be promoted either by repulsive steric interactions of the sugar headgroups or, most likely, by electrostatic interactions derived from adjacent carboxylate groups. At pH< pKa, when gels form, the carboxylate form of G-C18:0 is less abundant with respect to its carboxylic acid form; in the hypothesis of a symmetrical (head-tail-head-tail…[65]) arrangement and statistical distribution of the bolaform G-C18:0 molecule across the membrane, one does not reasonably expect a local accumulation of charge or of glucose headgroup. Bending should on the contrary be favoured for an antisymmetrical (head-head, tail-tail[65]) arrangement of the bolaform lipids not uniformly distributed across the membrane. Under these circumstances, one expects dense patches of glucose on one side of the membrane and COOH/COO⁻ on the other side. If the amount of COOH and COO⁻ groups is comparable within the patches, the surface area per molecule is the smallest[66] and the steric repulsion between the sugar headgroups could promote membrane bending. However, it is more reasonable to suppose that bending is driven by electrostatic repulsion, known to provide the largest surface area per lipid in carboxylate monolayers.[66] In this case, one must assume a higher concentration of COO$^-$ groups in the patches, as illustrated in **Figure 9e**. Finally, since all experiments are performed at T< T$_m$, the lipid mobility in the membranes is reduced; this is a condition that favours the stability of high curvature regions and, consequently, of defects, on time scales longer than the typical time scale of oscillatory rheology (above 10$^{-2}$ rad.s$^{-1}$). This hypothesis is supported by the fact that increasing the temperature above the T$_m$ promotes the formation of a viscous fluid, where defects are partially annealed.

The macroscopic elastic properties are resulting from a combination of defects at the nano/meso scale and the connections between the lamellar domains at the micrometer scale. If pH and temperature respectively play key roles in terms of total number of charge and its distribution in the membrane, we also stress the fact that the elastic properties are promoted, at a given pH, by addition of salt, of which the exact role is at the moment not fully clarified, yet. In fact, salt could have an antagonistic role: it could promote curing of the defects by neutralizing the negative charges from the COO$^-$ groups, but it could also reduce the elastic energy of the membrane,[67] thus promoting the formation of curved layer and, consequently, increasing the defect density. At the same time, in the case of a non-homogeneous distribution of salt on the membrane surface,[35] possibly due to an inhomogeneous distribution of the carboxylate groups, bridging of neighbor membranes and induction of line defects could also be a possibility. Interestingly, we also note that salt addition improves the elastic properties, thus suggesting its role on defect formation, but at the same time it promotes narrower d$_{(100)}$ and appearance of d$_{(200)}$ reflections in both SAXS and SANS profiles. Further work will be needed to better understand this point.

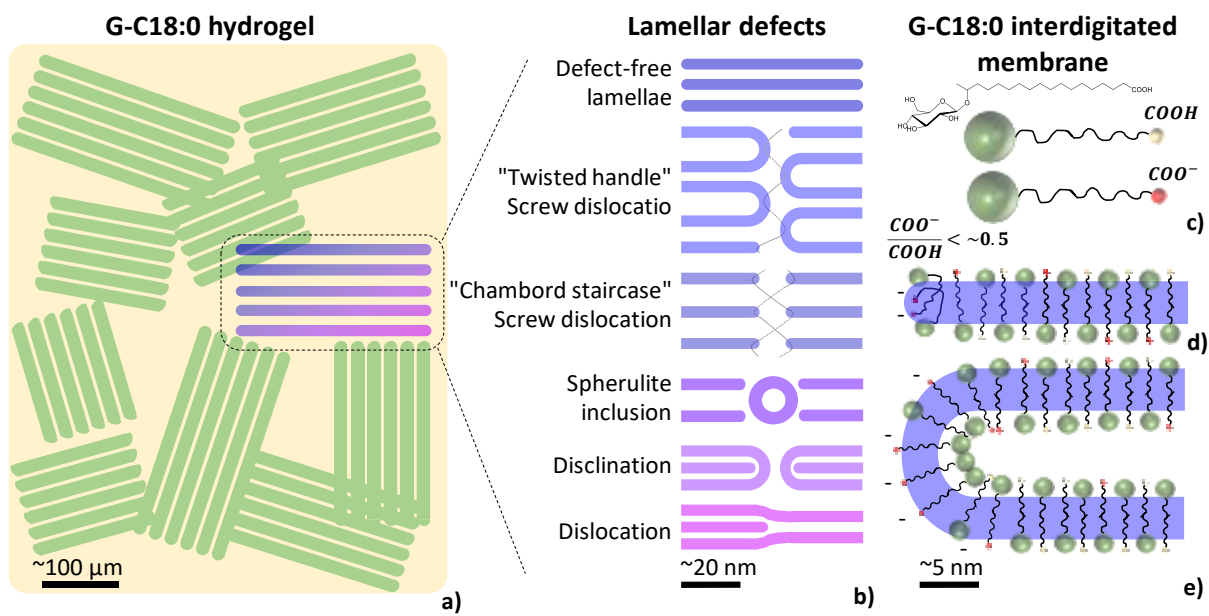

**Figure 9** – Hypothetical structure of the G-C18:0 hydrogel at T< $T_m$. a) Large lamellar domains (tens to hundred of microns) interconnect in water and b) each domain is composed of a highly defectuous lamellar phase, with typical defects possibly being displocations, disclinations, spherulite inclusions as well as screw dislocations.[12,41,62,63] c) A mixture of neutral and negatively-charged G-C18:0 molecules compose the interdigitated membranes in the gel. d) End-caps can be stabilized if negative charges accumulating at the periphery of the membrane while e) bending is rather promoted if patches of negatively-charged G-C18:0 molecules in an antisymmetrical (head-head, tail-tail) arrangement occur in the membrane.

**Conclusion**

This work shows that a new glucolipid obtained in large amounts from glucose and vegetable oils from the microbial fermentation of the modified yeast *S. bombicola ΔugtB1* is able to form lamellar rheo-thinning hydrogels in water in the pH range between 5 and 7.5 and ionic strength between 10 mM and 500 mM. Hydrogels are composed of a biphasic medium containing water and large interconnected domains (100-500 μm) of a kinetically-trapped lamellar phase stabilized by electrostatic interactions. The mechanical properties depend on both the high density of defects, as expected for lamellar hydrogels, but also on the multiscale structure, where lamellar domains of several hundred microns coexist with domains of few microns and disordered spherulitic inclusions, composed of entangled lamellar sheets of thickness less than 500 nm. The G-C18:0 glucolipid has a bolaform structure containing an ionizable COOH group. At pH below the pKa (8.4) the amount of negative charges is diluted enough to stabilize a lamellar structure composed of interdigitated GC18:0 molecules within the membranes having a thickness of 3.6 nm. The pH then defines an average charge density responsible for d-spacings in the order of 20 nm. Temperature, set below $T_m$ (37°C), freezes the G-C18:0 molecules in a kinetically-trapped state, thus having a strong impact on the actual charge distribution, which controls the formation of end-caps (**Figure 9d**) and membrane bending (**Figure 9e**), both phenomena being responsible for defect generation in the membrane (**Figure 9b**). Finally, ionic strength below ~500 mM consolidates the hydrogel, although its actual role at the mesoscale level impacting the macroscopic elastic properties of the gel is still unclear. This work shows the potential of new biobased compounds with possible applications in medicine, tissue engineering or cosmetics. At the same time, the unusual mechanical properties of single-lipid lamellar hydrogels should stimulate research in the field of complex fluids and biophysics for this new class of molecules.

**Acknowledgements:** Experimental and technical assistance: Dr. Andrea Lassenberger (SANS, ILL, Grenoble, France), Mr. Abdoul Aziz Ba (Sorbonne Université, Paris, France), Mrs Chloé Seyrig (Sorbonne Université, Paris, France), Dr. Patrick Davidson (PLM, LPS, Université Paris Sud, Orsay, France). Helpful discussions (lamellar phase and hydrogels, rheology): Dr. Jack Leng (LOF-CNRS, Bordeaux, France), Prof. Ulf Olsson (University of Lund, Lund, Sweden), Dr. Patricia Bassereau (Institut Curie, France), Frédéric Nallet (University of Bordeaux, Bordeaux, France), Dr. Bruno Demé (ILL, Grenoble, France), Dr. Thibaut Divoux (Massachusetts Institute of Technology, Cambrigde, MA, USA), Dr. Lionel Porcar (ILL, Grenoble, France), Dr. O. Diat (ICSM, Marcoule, France). J. L. Kaplan and A. Bonfanti (Engineering department, University of Cambridge) are acknowledged for assistance in using RHEOS software for the fitting procedure of the rheological data.

**Funding**: European Community's Seventh Framework Programme (FP7/2007–2013) under Grant Agreement No. Biosurfing/289219; European Synchrotron Radiation Facility (ESRF), Grenoble, France, under the experiment number SC 4778; Institut Laue Langevin (ILL), Grenoble, France, under the experiment number 9-13-778, DOI: 10.5291/ILL-DATA.9-13-778. S. R. and W. S. kindly acknowledge the VLAIO (Agentschap Innoveren & Ondernemen, Flanders Region, Belgium), Intercluster Vis Project AppliSurf: HBC.2017.0704.

**Author contributions:** GBM and NB performed the hydrogel experiments, analyzed the data and wrote the manuscript. PG performed microscopy experiments. SP provided assistance on the SANS experiments. DHM provided assistance on the SAXS experiments. WS and SR synthesized the glucolipid. CVS performed the hydrogenation experiments of the glucolipid.

**Competing Interests:** All authors declare no conflicts of interests.

**Supplementary Materials:**

Figures S1 to S5, Video 1-3

# Supplementary Information for

Single-molecule Lamellar Hydrogels from Bolaform Microbial Glucolipids

Ghazi Ben Messaoud, Patrick Le Griel, Sylvain Prévost, Daniel Hermida-Merino, Wim Soetaert, Sophie L. K. W. Roelants, Christian V. Stevens, Niki Baccile*

Correspondence to: niki.baccile@sorbonne-universite.fr

**This PDF file includes:**

Figs. S1 to S6

References for SI reference citations

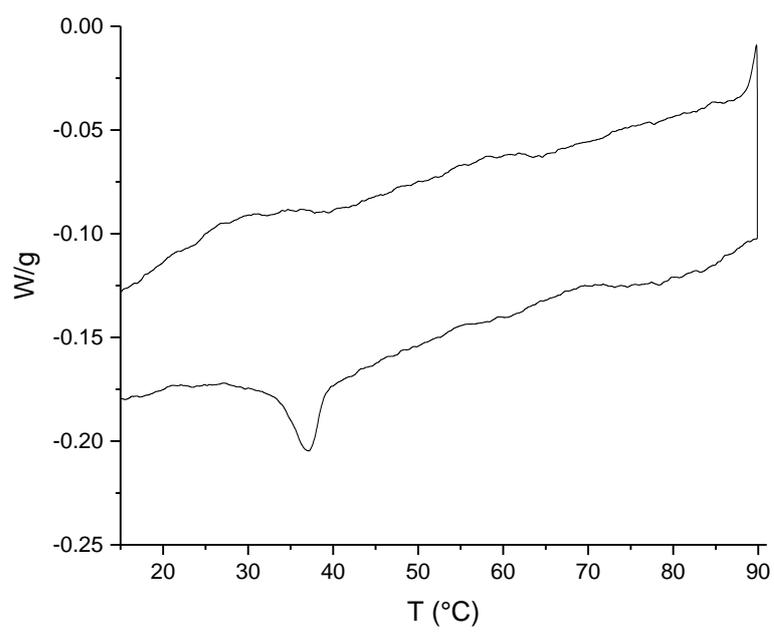

**Figure S 1 – Differential Scanning Calorimetry (DSC) thermogram of the native G-C18:0 powder acquired at 1°C/min**

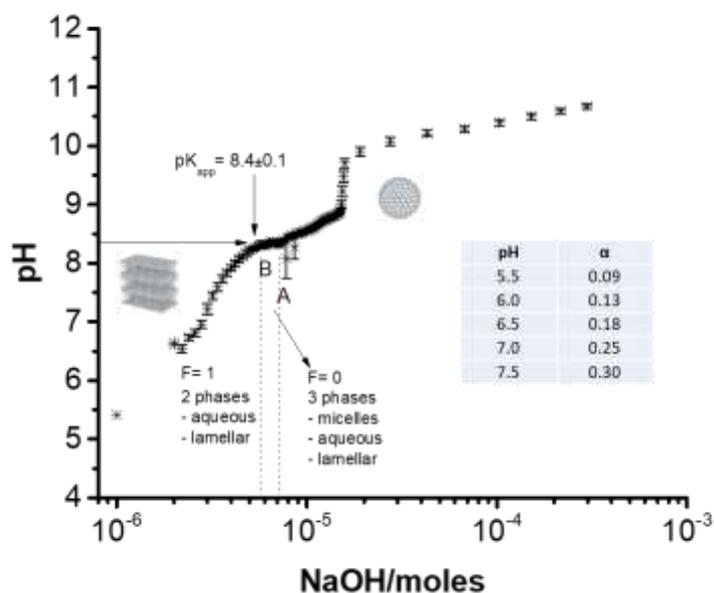

**Figure S 2** – Experimental determination of the apparent pKa of glucolipid G-C18:0 by titration using NaOH in water at T= 25°C (titrated volume: 1.8 mL). The ionization degrees are estimated using the Gibb's free energy rule applied to lamellar phases composed of fatty acids[1]

Estimating the ionization degree, α, for fatty acids as a function of pH is a long-date challenge which faces the problem of liquid-solid transition and coexistence of multiple phases. The problem of calculating α is in fact the problem of estimating the pKa. For stearic acid (which composes the fatty backbone of G-C18:0), the molecular pKa measured in organic solvents is between 4.9 and 5,[2] but this value varies between 5 and 7.6 when stearic acid is contained in a micellar environment in water.[3] In this work, we use the method of Cistola et al.[1], who applied the Gibb's phase rule to the titration curve of selected fatty acids below and above the melting temperature of the aliphatic chain. The advantage of this approach is multiple: 1) it is simple; 2) it has been used on the micelle-to-lamellar phase transitions of fatty acids in water; 3) it can be applied on our own experimental data (the titration curve) on a similar system. For the detailed description of the method, one can refer to ref. [1]. In this work, the region between points A and B in Figure S 2 satisfies the condition of invariance (F= 0), where composition is fixed, while the region below B (pH= 8.3) is characterized by one degree of freedom (F= 1), where composition of the lamellar phase can vary. α is estimated between pH 5 and pH 8.3, where α= 0.5 at B.

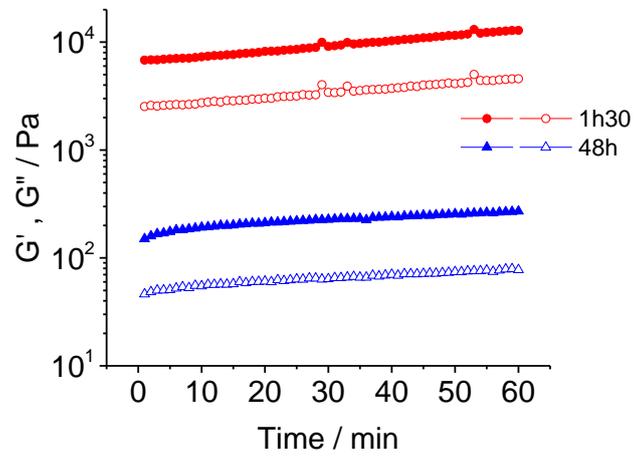

**Figure S 3** – Time evolution of G' (full symbols) and G'' (open symbols) (ω= 6.28 rad.s$^{-1}$ and γ= 0.05 %) for G-C18:0 ($C_{G-C18:0}$ = 5 wt%, pH= 6.7, [NaCl]= 167 mM) after 1h30 and 48 h from thermal annealing. Plate-plate geometry (25 mm), imposed normal force (NF = 0 N) and initial gap (0.5 mm).

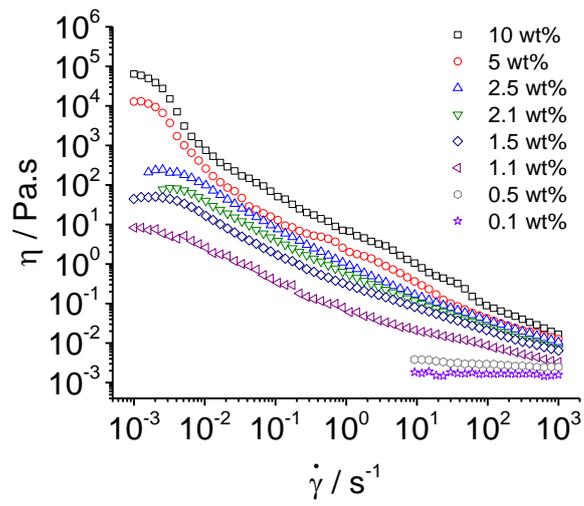

**Figure S 4 – Shear thinning profiles showing the evolution of viscosity with shear rate at different G-C18:0 concentrations at pH= 6.7 ± 0.1 and [NaCl]= 163 mM. Plate-plate geometry (25 mm) and an imposed gap of 0.5 mm are used**

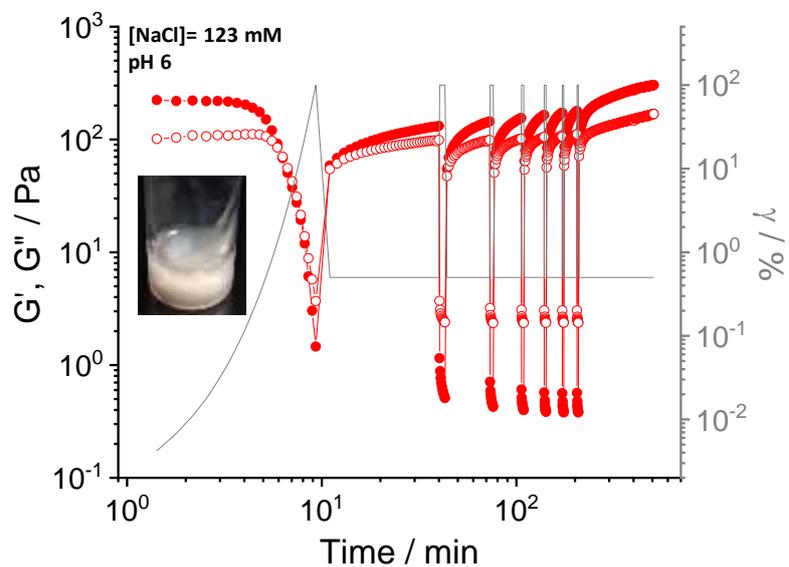

**Figure S 5** – Time evolution of G' (full symbols) and G'' (empty symbols) with the imposed shear strain (γ) at angular frequency (ω= 6.28 rad·s$^{-1}$). Logarithmic increase of the shear strain (4·10$^{-3}$ < γ < 100 %) during 10 min followed by a recovery at γ = 0.5 % (upper limit of the LVER) during 30 min, followed by 6 cycles of step strain experiments (γ = 100% during 2 min followed by γ = 0.5 % during 30 min (during the first 5 cycles) and 300 min in the last cycle. Plate-plate geometry (25 mm) and an imposed normal force (NF = 0 N) with an initial gap (0.5 mm) are used.

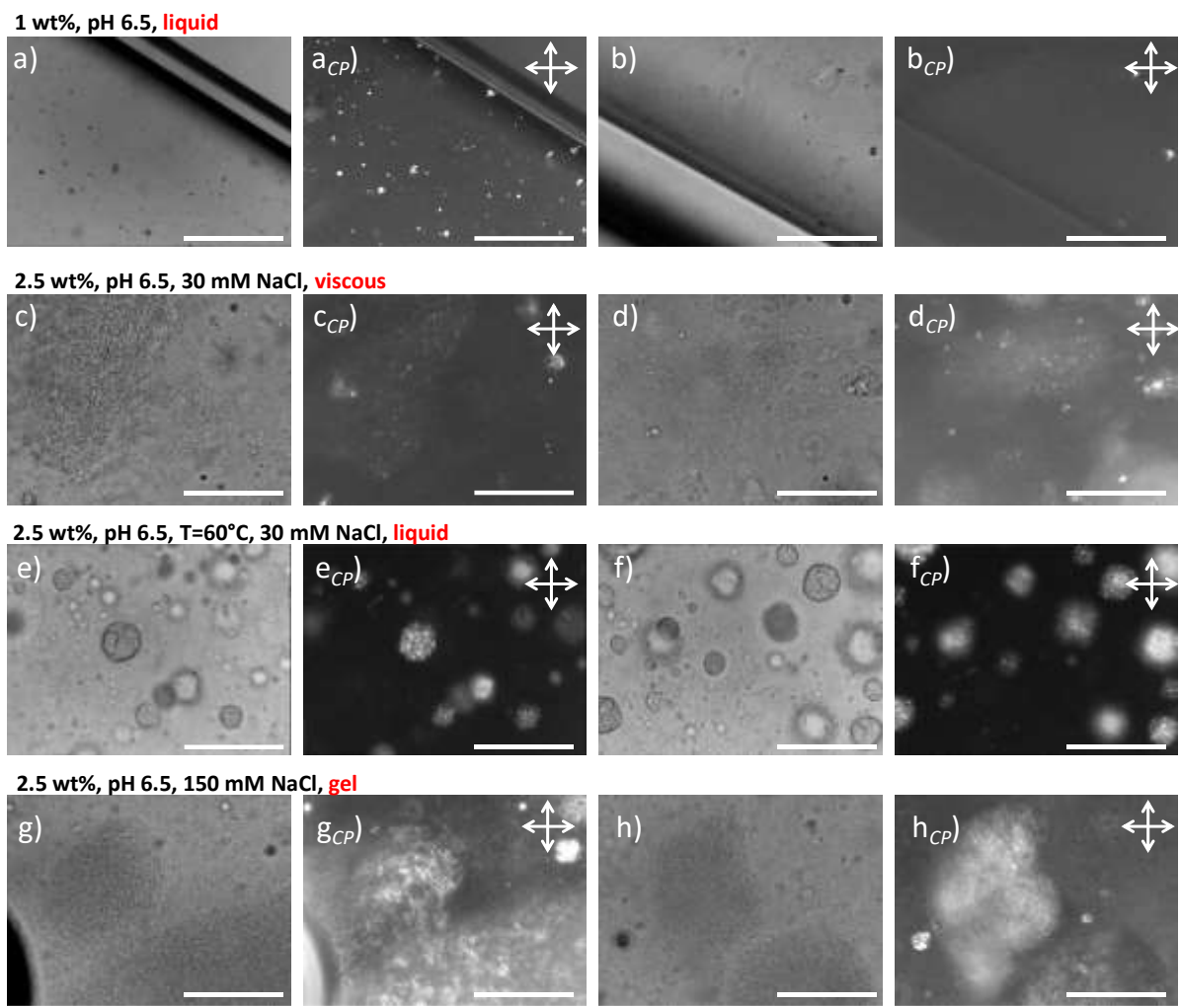

**Figure S 6** – Bright Field and polarized light microscopy (*CP*: Crossed Polarizers) images of a set of G-C18:0 samples at pH 6.5 in water (scale bar: 100 µm). Samples are prepared in flame-sealed flat capillary of 200 µm thickness, as in ref. [4]. Exact conditions are given on top of each series of images. All images are recorded after 24 h from preparation. All samples are kept at room temperature ~ 23°C, except in images e)-f), in which samples are kept at T= 60°C. The physical state of the sample are explained in red for each series of images. Images in c)-d) and e)-f) refer to the same sample, although in e)-f) the sample has been placed at T= 60°C.

*Considerations on the lamellar phase*

Nematic phases are characterized by orientational order and loss in translational order, and the latter could be promoted by the development of defects, like screw dislocations. Dhez *et al.* have shown the effect of dislocation defects on the diffraction profile of a lamellar phase close to the lamellar-to-nematic transition in lipid-surfactant-water systems, pointing at the complexity of a straightforward attribution of SAXS profiles under these conditions.[5] SAXS is commonly employed to characterize the lamellar nature of a lipid phase, however, the diffraction peak alone in the SAXS/SANS data in the G-C18:0 hydrogels is often very broad and it may not unambiguously help to discriminate between nematic order and a defectuous lamellar phase,[5,6] especially for the viscous solutions at low ionic strength lacking of a second order peak. The $d_{(100)}$ peak is much broader (Figure 2 in the main text) than what one classically finds in lamellar phases, and it could be interpreted as a coagel-to-gel transition[7] or to a nematic phase.[6] The former case is excluded, because the characteristic fibrillar crystals are never observed and gel always forms below the lipid $T_m$. Concerning the presence of a nematic phase, the SAXS data of hydrogels (e.g., pH 7, [NaCl]≤ 250 mM, Figure 2d in the main text) can be fitted with a lamellar form factor, while the diffuse scattering peak below 1 nm$^{-1}$ can be fitted with a lamellar structure factor taking into account displacement fluctuations about the ideal lattice position.[8] Analogous SAXS profiles are also reported for biomembranes.[9] Meanwhile, all cryo-TEM data in our possession show the systematic presence of flat sheets being "infinite" in the planar dimension and polarized light microscopy data never display any typical texture of nematic order but they all rather closely look like the whispy textures found in lamellar hydrogels, as commented in the main text.